\documentclass{iaus}
\usepackage{graphicx}

\title[Rotational Spectroscopy of PAHs] %% give here short title %%
{Rotational Spectroscopy of PAHs: Acenaphthene, Acenaphthylene and Fluorene}

\author[Thorwirth et al.]   %% give here short author list %%
{S. Thorwirth%
  \thanks{Present address: Max-Planck-Institut f\"ur Radioastronomie, Auf dem H\"ugel 69,
  53121 Bonn, Germany},
  P. Theul{\'e}\thanks{Present address: Physique des interactions ioniques et mol\'{e}culaires,
  Universit\'{e} de Provence, Centre de Saint J\'{e}r\^{o}me, 13397 Marseille Cedex 20, France}, 
  C. A. Gottlieb, \break M. C. McCarthy, \and P. Thaddeus}

\affiliation{Department of Engineering and Applied Sciences, Harvard University,\break Pierce Hall, 29 Oxford Street, Cambridge, MA 02138, U.S.A. and \break Harvard-Smithsonian Center for Astrophysics,
\break 60 Garden Street, Cambridge, MA 02138, U.S.A. 
\break email: sthorwirth@mpifr-bonn.mpg.de
}

\pubyear{2005}
\volume{231}  %% insert here IAU Symposium No.
\pagerange{1--2}
\date{?? and in revised form ??}
\jname{Astrochemistry - Recent Successes and Current Challenges}
\editors{D.~C. Lis, G.~A. Blake \& E. Herbst, eds.}
\begin{document}

%%%%% please include in the following indexes:
%
% SUBJECT INDEX: laboratory data, large molecules, PAH
% 
% MOLECULE INDEX: C12H10, C12H8, C13H10
%
%%%%%%%%%%% 

\maketitle

\begin{abstract}
Pure rotational spectra of three polycyclic aromatic hydrocarbons -- acenaphthene, acenaphthylene and fluorene -- have been obtained by Fourier transform microwave spectroscopy of a molecular beam 
and subsequently by millimeter wave absorption spectroscopy for acenaphthene and fluorene.
The data presented here will be useful for deep radio astronomical searches for PAHs employing large radio telecopes.
\keywords{astrochemistry, molecular data, ISM: molecules}
\end{abstract}


Polycyclic aromatic hydrocarbons (PAHs) have been studied extensively in the laboratory over the last
20 years (see \cite{salama1999} and \cite{tielens2004} for reviews) owing to their astronomical significance as possible carriers of the unidentified infrared bands (UIRs, \cite[e.g. Allamandola et al. 1989]{allamandola1989}). These studies have been performed almost exclusively in the uv-, optical-, and infrared regions of the electromagnetic spectrum. Very little is known, however, about the
rotational spectra of small and polar PAHs, since microwave studies have been reported so far only
for azulene (C$_{10}$H$_8$, \cite{huber2005} and references therein) and corannulene (C$_{20}$H$_{10}$, \cite[Lovas et al. 2005]{lovas2005}).

In the present study, we have investigated the rotational spectra of selected small PAHs (see Fig. 1)
employing Fourier transform microwave (FTM) spectroscopy (\cite[Balle \& Flygare 1981]{balle1981}) using the spectrometer at Harvard (\cite[McCarthy et al. 1997, 2000]{mccarhty1997,mccarthy2000})
in combination with a heated nozzle recently developed for studies of low volatile compounds 
(\cite[Thorwirth et al. 2005]{thorwirth2005}).
\begin{figure}[h]
\begin{center}
\includegraphics[width=11cm]{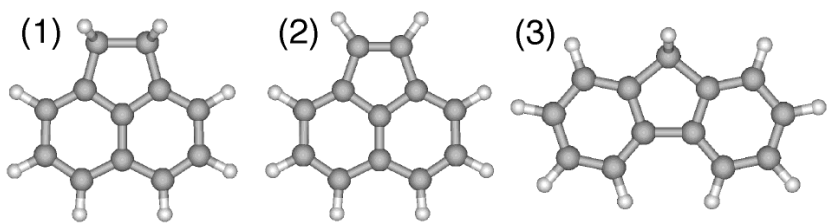}
\caption{Molecular structures of acenaphthene (C$_{12}$H$_{10}$, {\bf 1}), acenaphthylene (C$_{12}$H$_{8}$, {\bf 2}) and
fluorene (C$_{13}$H$_{10}$, {\bf 3}).}
\end{center}
\end{figure}
Initial searches were guided by rotational constants obtained from quantum chemical calculations performed at the B3LYP/cc-pVTZ level of theory (see Table 1) using the program package Gaussian03 (\cite[Frisch et al. 2003]{frisch2003}). All three molecules exhibit $b$-type rotational spectra 
\begin{table}[h!]
\caption{Rotational constants (in MHz) and dipole moments $\mu$ (in D) for ({\bf 1}), ({\bf 2}) and ({\bf 3}) as determined in the present study. }
\begin{center}
\begin{tabular}{lrrrr|rrr} \hline 
         & \multicolumn{4}{c}{B3LYP/cc-pVTZ} &  \multicolumn{3}{c}{Experiment} \\
Molecule & $A_e$ & $B_e$ & $C_e$ & $\mu$ &  $A_0$ & $B_0$ & $C_0$   \\ \hline
Acenaphthene ({\bf 1})   & 1416.5 & 1200.6 & 655.0 & 0.9 & 1410.3 & 1193.9 & 652.1 \\
Acenaphthylene ({\bf 2}) & 1520.7 & 1228.2 & 679.4 & 0.3 & 1511.8 & 1220.6 & 675.5 \\
Fluorene ({\bf 3})       & 2195.1 &  588.2 & 465.2 & 0.5 & 2176.2 &  586.7 & 463.6 \\ \hline
\end{tabular}
\end{center}
\end{table}
and are calculated to be moderately polar, with dipole moments of order 0.3 to 0.9 D. Rotational transitions were found readily for all three molecules and based on improved predictions several tens of lines could be measured for each one. Rotational constants obtained from least-squares analyses of the experimental data are shown in Table 1. As can be seen, the calculated equilibrium values
and experimentally obtained ground state rotational constants agree very well, to within 1\%. Subsequently, selected rotational transitions of acenaphthene and fluorene could also be measured by standard millimeter wave absorption spectroscopy at 90\,GHz.

The present investigation highlights the potential of FTM spectroscopy for the characterization of polar PAHs, including the nitrogen variants (PANHs; e.g., see the contribution by D. Hudgins, these proceedings).

A detailed account on the present study will be given elsewhere.

\acknowledgments{This work was supported in part by NASA grant NAG5-9379 and NSF grant CHE-0353693.
S.~Thorwirth is grateful to the Alexander von Humboldt-Foundation for a Feodor Lynen
research fellowship. P. Theul{\'e} would like to thank the Swiss National Science Foundation
for a research fellowship.}

%\appendix

%\begin{discussion}

%\discuss{Massey}{I'm wondering if you have considered the expected intrinsic dispersion in absolute
%magnitude of WRs --— if you consider the (large) mass range that becomes an
%early WN or late WC according to the evolutionary models, wouldn't you expect a large
%dispersion in M$_v$?}

%\discuss{van der Hucht}{Wonderful.}


\begin{thebibliography}{}

\bibitem[Allamandola et al. (1989)]{allamandola1989} Allamandola, L.~J., Tielens, A.~G.~G.~M., 
\& Barker, J. R. 1989, \textit{Astrophys. J. Suppl. S.}, 71, 733

\bibitem[Balle and Flygare (1981)]{balle1981} Balle, T. J. \& Flygare, W. H. 1981, \textit{Rev. Sci. Instrum.}, 51, 33

\bibitem[Frisch et al. (2003)]{frisch2003} Frisch, M. J., Trucks, G. W., Schlegel, H. B. \textit{et al.} 2003, Gaussian03, Revision B.04, Gaussian, Inc., Wallingford CT

\bibitem[Huber et al. (2005)]{huber2005} Huber, S., Grassi, G., \& Bauder A. 2005, \textit{Mol. Phys.},
103, 1395

\bibitem[Lovas et al. (2005)]{lovas2005} Lovas, F. J., McMahon, R. J., Grabow, J.-U., Schnell,
M., Mack, J., Scott, L. T., \& Kuczkowski, R. L. 2005, \textit{J. Am. Chem. Soc.}, 127, 4345 

\bibitem[McCarthy et al. (1997)]{mccarhty1997} McCarthy, M. C., Travers, M. J., Kov{\'a}cs, A., Gottlieb, C. A., \& Thaddeus, P. 1997, \textit{Astrophys. J. Suppl. S.}, 113, 105

\bibitem[McCarthy et al. (2000)]{mccarthy2000} McCarthy, M. C., Chen, W., Travers, M. J., \& Thaddeus, P. 2000, \textit{Astrophys. J. Suppl. S.}, 129, 611

\bibitem[Salama (1999)]{salama1999} Salama, F. in \textit{Solid Interstellar Matter: The ISO Revolution} (Eds.: L. d'Hendecourt, C. Joblin, A. Jones), Springer-Verlag, New York, 1999, 65

\bibitem[Thorwirth et al. (2005)]{thorwirth2005} Thorwirth, S., McCarthy, M.~C., Gottlieb, C.~A.,
 Thaddeus, P., Gupta, H., \& Stanton, J.~F. 2005, \textit{J. Chem. Phys.}, 123, 054326

\bibitem[Tielens \& Peeters (2004)]{tielens2004} Tielens, A.~G.~G.~M. \& Peeters, E. in {\it The Dense Interstellar Medium in Galaxies} (Eds.: S.~Pfalzner, C. Kramer, C. Staubmeier, and A. Heithausen) Springer proceedings in physics, Vol. 91, Springer-Verlag, Berlin, 2004, 497



\end{thebibliography}
\end{document}